\documentstyle[epsfig, subfigure]{elsart}

\def\compile2{Elsepaper.tex
rm Elsepaper.bbl Elsepaper.aux Elsepaper.ttt Elsepaper.fff
rm Elsepaper.lof Elsepaper.log
latex Elsepaper
bibtex Elsepaper
latex Elsepaper
latex Elsepaper
dvips  -o Elsepaper.ps Elsepaper

ghostview Elsepaper.ps&

}

\begin{document}
\begin{frontmatter}

\title {Spatiotemporal bifurcations in plasma drift-waves}
\author {Alex Madon}
\address {CNRS, Centre de Physique Th\'eorique, Luminy,
         Case 907, F-13288 Marseille, France,
         madon@cpt.univ-mrs.fr}
\author{Thomas Klinger}
\address{Institut f\"ur Experimentalphysik,
Christian-Albrechts-Universit\"at,
Kiel, Olshausenstrasse 40-60, D-24098 Kiel, Germany}

\begin{abstract}
Experimental data from an experiment on drift--waves in plasma
is presented. The experiment provides a space--time diagnostic and
has a control parameter that permits the study of the transition
from a stable plasma to a turbulent plasma.
The biorthogonal decomposition is used to analyse the data.
We introduce the notion of complex modulation for two--dimensional
systems. We decompose the real physical system into complex
modulated monochromatic travelling waves and give a simple model
describing the speed doubling observed in the data as the control
parameter increases.
\end{abstract}

\begin{keyword}turbulence , drift-waves,
bifurcation,
PACS : 52.35.R, 52.35.K, 47.20.K.
\end{keyword}

\end{frontmatter}

\section{introduction}
%%%%%%%%%%%%%%%%%%

The stability of spatially extended dynamical systems and the transition
towards statio-temporal chaos is still a wide open problem. Considerable
progress has been made recently by models for pattern formation
\cite{hohenberg93} and advanced three-wave interaction schemes
\cite{chow95,kaup79}. The plasma turbulence phenomenon \cite{tsytovich}
is considered to be
one possible application to the concept of spatio-temporal chaos.
In particular, drift wave turbulence is of
special interest for the understanding of anomalous transport
in magnetically confined plasmas \cite{horton84}. For the investigation
of the transition from a stable plasma to drift wave turbulence,
the study of nonlinear model
descriptions is unambiguous\cite{horton90}.

In the present paper, we present a simple dynamical model for
spatio-temporal bifurcations towards chaos. It is obtained from
the numerical analysis of new experimental observations of the
space-time structure of current-driven drift waves in cylindrical
geometry.

In Section \ref{exp}, drift waves are briefly introduced.
The experimental arrangement as well as the diagnostic tools for
the measurement of the spatio-temporal structure of regular and
turbulent drift waves are described. For the present investigation,
three paradigmatic data sets
are considered for three different values of the control parameter :
one single propagating mode, nonlinear interaction
of drift modes, and fully developed turbulence. These different
dynamical states result after successive bifurcations caused by
an increase of an appropriately chosen control parameter.
The powerful tool
of biorthogonal decomposition \cite{LimaCom,dudok94} is used to
analyse the experimental data. The results of this analysis provide
the basis for the model description.
In Section \ref{moduldef}, the complex modulations of
two--dimensional spatio--temporal systems are defined and
the importance of modulated monochromatic travelling waves
(modulated MTW) is underlined.
The modulated MTWs are identified in Section \ref{LookModTW}
and characterized in Section \ref{secorg}. A speed doubling
phenomenon is observed. A low--dimensional model is provided
in Section \ref{secmodel}.
\section{Experimental investigations}\label{exp}
Drift waves are universal instabilities of magnetically confined plasmas.
In the present study we consider a cylindrical plasma
column with a constant
axial homogenous magnetic field $\vec{B}=B\vec{z}$ and a radial electron
density profile $n_{\rm e}(r)$. A fluid description \cite{nicholson} of
the plasma shows that the electrons drift in the azimuthal direction
$\vec{\theta}$ with diamagnetic drift velocity
\begin{equation}
  \vec{v}_{\rm dia}=-\frac{k_{\rm B}T_{e}}{eB}L^{-1}\,\vec{\theta}
\end{equation}
where $L^{-1}=\frac{1}{n(r)}\frac{{\d}n(r)}{{\d}r}$
describes the inverse gradient
length of the density profile. From the thermodynamic point of view, the
diamagnetic drift provides a source of free energy for the drift
instability \cite{krall}. The drift instability propagates as a plasma
wave in the azimuthal direction with diamagnetic velocity. The wave
number is predominantly in the $\vec{\theta}$-direction, but has
a small $\vec{z}$-component to allow the electrons to flow freely
along the magnetic field lines. The frequency of the instability
is roughly given by $\omega^*=(m/r_0)v_{\rm dia}$, where $r_0$ is
the position of the maximum density gradient and $m$ is the azimuthal
mode number. The linear analysis of a cylindrical, weakly ionized plasma
has revealed that the growth rate of drift instabilities is strongly
enhanced by electron neutral collisions as well as by an additional
electron drift along the $\vec{z}$-axis \cite{ellis80}.
Generally, only the onset regime of
growing drift instability allows a linear description.
In the nonlinear regime, a large variety of new dynamical phenomena
become important (for a detailed review see
Refs.~\cite{horton84,horton90}).
Of particular interest is the transition from a stable plasma to
turbulence, because fluctuation-induced
transport plays a decisive role
in the plasma edge physics of fusion devices \cite{carreras92,wagner93}.

The experimental investigations are carried out in the central section
of a large magnetized triple plasma device. A schematic drawing is
shown in Fig.~\ref{kiwi}.
\begin{figure}[htb]
 \centering
% \epsfig{file={kiwi.ps},width=\textwidth,angle=90}

%bbllx=2.5cm,bblly=19.5cm,bburx=18.5cm,bbury=27.5cm}
 \caption{Schematic representation of the large triple plasma device.
          The plasma is produced in two independent source chambers.
          Drift waves are observed in the plasma column in the
          magnetized midsection.}
 \label{kiwi}
\end{figure}
A homogenous quiet plasma is produced in two independent
multidipole-confined discharge chambers \cite{limpaecher73}.
For the present studies, argon is used as the filling
gas and the degree of ionization
is below $0.1\%$. The magnetized central section is
separated from the plasma chambers by two electrically isolated mesh
grids (transparency $50\%$). The plasma produced in the discharge
chambers enters the central section by (i) diffusion and (ii) drift.
The diffusion process (i) is based on the axial density gradient
between the plasma chamber and the central section. The effect of
magnetic mapping by the fringing magnetic field lines at the end of the
central section is minimized by a particular magnetic
compensation technique
\cite{pierre86}. An additional electron drift (ii) is superimposed by
biasing the grid positively with respect to the plasma potential in
the source chamber. This electron drift is known to destabilize
drift waves (see above) and is consequently an appropriate control
parameter for the dynamics of the system. Experimentally, the axial
electron drift is varied by the bias of the injection grid
$U_{\rm G1}$ (cf.~Fig.~\ref{kiwi}). The second grid is considered
as a plasma loss surface and remains at anode potential, i.e :
 $U_{G2}=0$.
In order to be close to the threshold value for the onset of the drift
instability, only one plasma chamber is operated and a gradient--driven
electron drift is present. The most important discharge parameters and
the plasma parameters for the present measurements are summarized in
Table \ref{param}. The radial profiles of the electron-density, plasma
potential, and electron-temperature are plotted in Fig.~
\ref{profiles}. A more detailed description of the experiment can
be found in Ref.~\cite{latten95}.
\begin{table}[htb]
 \begin{center}
  \caption{List of discharge parameters and plasma parameters for which
          the experimental investagation is performed.}
  \label{param}
  \begin{tabular}{l|l|c}
     parameter                    & symbol & value\\
     \hline\hline
     magnetic field               & $B$           & $70{\rm mT}$ \\
     neutral gas pressure (argon) & $p$           & $5.6\cdot 10^{-2}
{\rm Pa}$ \\
     discharge voltage            & $U_{\rm d}$   & $60{\rm V}$\\
     discharge current            & $I_{\rm d}$   & $13{\rm A}$\\
     plasma radius (FWHM)         & $R$           & $0.1{\rm m}$\\
     plasma length                & $L$           & $1.8{\rm m}$\\
     $e$ drift velocity           & $v_{\rm d,e}$ &
$\leq 0.4v_{\rm th,e}$\\
     \hline
     electron density (center)     & $n_{\rm e}$   &
$5 \cdot 10^10{\rm cm}^{-3}$\\
     electron temperature (center) & $T_{\rm e}$   & $1.5{\rm eV}$\\
     ion temperature (center)      & $T_{\rm i}$   & $0.03{\rm eV}$\\
     \hline\hline
  \end{tabular}
 \end{center}
\end{table}
\begin{figure}[htb]
 \centering
 \caption{Radial profiles of a) the electron density, b) the plasma
potential, c) the electron temperature}
 \label{profiles}
\end{figure}
The spatiotemporal structure of regular and turbulent drift waves is
investigated by an azimuthally arranged multi-channel Langmuir probe
array \cite{latten95}. Each single Langmuir probe provides the temporal
fluctuations of the plasma density. For the present experiment, a
circular arrangement of $N=64$ probes at constant radial and axial
position is used.
%*******************
The fixed radial position of the probes, imposed by technical
constraints, doesn't allow to address here the interesting problem
of the radial profile of the perturbation.
%*******************
The probe array provides the temporal evolution of the spatial
structure of drift waves which propagate mainly in the azimuthal
direction. The temporal resolution is given by the maximum sample
rate of the aquisition system ($\Delta t=1\mu{\rm s}$), and the spatial
resolution is given by the azimuthal angle between each two probes
($\Delta x=2\pi/64$). The data is stored as an $N \times M$-matrix
$u_{i,j}=n_{e}(i\Delta x,j\Delta t)$ where $N=64$ (space) and $M=2048$
(time).
%*************************
Note that investigations of spatiotemporal phenomena
can be done with one probe by a method based on conditional averaging
\cite{Iizuka}.
%*************************

In order to study the bifurcation behaviour of drift waves,
the data sets of three representative dynamical states are considered:
${\cal U}_1$ a single monochromatic drift mode, ${\cal U}_2$ drift
modes with nonlinear interaction, ${\cal U}_3$ turbulent drift waves.
 The
different dynamical states ${\cal U}_k$ are recorded successively by
increasing the accessable control parameter $U_{\rm G1}$.
In Fig.~\ref{data},
the three data sets ${\cal U}_k$ are shown as greylevel plots.
\begin{figure}[htb]
 \caption{Greylevel plots of the three drift wave data sets. Grey
corresponds
          to the zero level, black is the minimum, and white is the
 maximum
          deviation from the equlibrium electron density in the plasma.
The
          vertical axis is labelled with the probe number index, the
          horizontal axis is the time in multiples of
$\Delta t=1\mu{\rm s}$.
          Control parameter values are (a) $U_{G1}=4.5{\rm V}$, (b)
          $U_{G1}=4.6{\rm V}$, (c) $U_{G1}=10{\rm V}$.}
 \label{data}
\end{figure}
Data set ${\cal U}_1$ [Fig.~\ref{data}(a)] is a single propagating mode,
data set ${\cal U}_2$ [Fig.~\ref{data}(b)] shows interacting drift modes,
and data set ${\cal U}_3$ [Fig.~\ref{data}(c)] is the state of strong
drift wave turbulence. These three data sets are the basis for the
following analysis and model description. For this purpose the plasma
is considered as a spatiotemporal dynamical system, whose state is
described by a function $u_{\epsilon}(x,t)$ where $\epsilon$ represents
the complete set of experimental parameters.
For the present investigations
only one parameter, the grid bias $U_{\rm G1}$, is varied.
%%%%%%%%%%%%%%%%*********
The grid bias has been chosen as control parameter
because it allows to consider
that all other parameters remain in good approximation constant.
%%%%%%%%*************

Each dynamical
state given by the data set ${\cal U}_k$ corresponds to a control
parameter
value $\epsilon_k$.

%
% add here the BOD analysis
%

\medskip
The analysis tool used here is the biorthogonal
decomposition (BOD). This tool provides a way to study in
 the space and time properties simultaneously.
We present here just the most important features of the BOD.
(For more details see Ref.~\cite{LimaCom})

%%%%%%%%%BOD
Suppose that our system is described by a function $u(x,t)$ defined
on a spatial range $X$ and a temporal interval $T$.
In the experimental situation, $X$ is the domain of the azimuthal
angle $x$ in cylindrical coordinates, i.e. $X=[0,2\pi]$.
The biorthogonal decomposition provides the smallest linear
subspace $\chi(X)$ containing the phase space trajectory
$\xi_t$ (described as time $t$ runs) defined by :

\begin{equation}
\forall x \in X,\ \xi_t(x)=u(x,t).
\end{equation}

The set of all the vectors $\xi_t$ is the trajectory
and the evolution of $\xi_t$ is the dynamics of the system.

\medskip

The biorthogonal decomposition also provides the smallest
linear subspace $\chi(T)$ containing the spatial structure $\xi_x$
 (described as the spatial position $x$ varies) defined by :

\begin{equation}
\forall t \in T,\ \xi_x(t)=u(x,t).
\end{equation}

In the present paper, the $L^2$ scalar product is used to
define $H(X)$ and $H(T)$, the Hilbert spaces of the
functions of $x$ defined on $X$, and
the functions of $t$ defined on $T$ respectively.
The BOD is the spectral analysis of the operator $U$,
which acts from $H(X)$ into $H(T)$ :

\begin{equation}
(U\phi)(t)=\int u(x,t)\phi(x) dx,
\end{equation}

where $U$ defines a one--to--one relation between the
vectors of $\chi(X)$ and $\chi(T)$, the orthogonal complements of the
kernels of $U$ and its adjoint $U^*$.

\medskip

If $U$ is compact, as it is here
the BOD decomposes $u(x,t)$ into temporal  and
spatial orthogonal modes and $u(x,t)$ can be written as follows :

\begin{equation}
u(x,t)=\sum \alpha_n \psi_n(t)\phi_n(x),
\end{equation}

with $\alpha_1\geq\alpha_2\geq\dots\geq 0$, and the orthogonality
 relations $(\phi_n,\phi_m)=\delta_{n,m}$ and
$(\psi_n,\psi_m)=\delta_{n,m}$. The $\phi_n$ are called topos,
 and the $\psi_n$ chronos.

\section{Modulated travelling waves}\label{moduldef}
%%%%%%%%%%%%%%%

In this section we discuss the conditions for
a two--dimensional system to be considered a modulated
travelling wave.
The introduction of modulated travelling waves is of interest
as we shall see in the drift wave turbulence, where
the decomposition of a high--dimensional system into a set of
two--dimensional sub--systems allows us to
focus on the dominating properties of the dynamical behaviour.
The way in which a monochromatic travelling wave may be deformed
in our case of study is now analysed in the following theorm for
which we need first to introduce some simple definitions concerning
a complex formalism for the corresponding modulations.

\medskip

Let us now specify what is meant by two--dimensional projections.
of a N-dimensional system be described by its BOD

\begin{equation}
u(x,t)=\sum_{k=1}^N a_k \psi_k(t)\phi_k(x).
\end{equation}

defined on the space interval
$X=[0,2\pi]$ and on the time interval $T=[t_0,t_1]$.

\begin{defn}
The {\bf projection} of a system $u(x,t)$ onto two vectors of index
$m$ and $n$ is the two dimensional system $u_{m,n}(x,t)$ described by :

\begin{equation}
u_{m,n}(x,t)=a_m \psi_m(t)\phi_m(x)+a_n \psi_n(t)\phi_n(x)
\end{equation}

\end{defn}

Thus the projection of the dynamics $\xi_t$
 (associated with the spatio--temporal system $u(x,t)$)
onto the eigenvectors of index $m$ and $n$ is simply
the projection of $\xi_t$ onto the two topos $\phi_m$ and $\phi_n$, i.e.

\begin{equation}
\forall x \in X,\  \xi_t^{m,n}(x)=u_{m,n}(x,t).
\end{equation}

The projection of the spatial structure (also given by the
spatio--temporal system $u(x,t)$) onto the eigenvectors of
index $m$ and $n$ is simply the projection of $\xi_x$
onto the two chronos $\psi_m$ and $\psi_n$.

\begin{equation}
\forall t \in T,\   \xi_x^{m,n}(t)=u_{m,n}(x,t)
\end{equation}

The schematic drawing Fig.\ref{bodsk} illustrates this projection
process of the dynamics and the spatial structure.

\begin{figure}
\centering
\caption{The operator $U$ maps  the subspace of $\chi(X)$ spanned by
 $\phi_1$ and $\phi_2$ into the subspace of $\chi(T)$ spanned by
$\psi_1$ and $\psi_2$.}
\label{bodsk}
\end{figure}

\begin{defn}
A {\bf modulatrix} ${\cal M}$ is a pair of complex valued continuous
functions $M(x)$ and $N(t)$. $M$ is called the spatial modulatrix,
 and $N$ the temporal modulatrix.
\end{defn}

Note that each complex--valued continuous function
is the parametric representation of an arc.
Using the vocabulary of complex analysis
(see for instance \cite{Henrici}), the arc $\Gamma_X$ is defined as

\begin{equation}
\Gamma_X\  :\  M=M(x), \ 0\leq x \leq 2\pi .
\end{equation}

\begin{defn}
A modulatrix ${\cal M}=(M,N)$ is a {\bf continuous phase modulatrix}
 if $M$ and $N$ can be written as

\begin{equation}
M(x)=A(x)e^{iF(x)}
\label{contx}
\end{equation}

and

\begin{equation}
N(t)=B(t)e^{iG(t)},
\label{contt}
\end{equation}

where $A$, $B$, $F$, and $G$ are continuous functions. It will be called
a {\bf regular continuous phase modulatrix} if moreover the increase
of the argument of $F$, $D_F=F(2\pi)-F(0)$, and
the increase of the argument of $G$, $D_G=G(t_1)-G(t_0)$, are both
equal to zero.
\end{defn}

It is known that a continuous complex--valued function
can be written as a
 product of a continuous modulus function and a continuous argument
 function if the function is never equal to 0 (see \cite{Henrici}).
The modulus function is unique and two different argument functions
differ by a constant integral multiple of $2\pi$.
The real numbers $D_F=F(2\pi)-F(0)$ and $D_G=G(t_1)-G(t_0)$
are independent of the choice
of the argument functions $F$ and $G$ \cite{Henrici}.

%%%%%%%%%%%%%%% COMPLEXIFICATIONS

\begin{defn}
The {\bf spatial complexification} of two--dimensional system $u(x,t)$
whose BOD is

\begin{equation}
u(x,t)=\alpha_1\psi_1(t)\phi_1(x)+\alpha_2\psi_2(t)\phi_2(x)
\label{bod2d}
\end{equation}

 is

\begin{equation}
Z(x)=\alpha_1\phi_1(x)+i\alpha_2\phi_2(x).
\end{equation}

The corresponding {\bf temporal complexification} is

\begin{equation}
Y(t)=\alpha_1\psi_1(t)+i\alpha_2\psi_2(t).
\end{equation}

\end{defn}

Note that the spatial complexification is a representation of the
spatial structure in the complex plane, and the temporal
complexification is a representation of the dynamics in
the complex plane.

\medskip

If the complexifications are never zero, the argument function
can be introduced by the following definitions :

\begin{defn}
A spatial complexification $Z(x)$ is a {\bf phase continuous
complexification} if it can be written as :

\begin{equation}
Z(x)=C(x)e^{iQ(x)}.
\end{equation}

\end{defn}

\begin{defn}
A temporal complexification $Y(t)$ is a {\bf phase continuous
complexification} if it can be written as :

\begin{equation}
Y(t)=D(t)e^{iR(t)}.
\end{equation}

\end{defn}

Of particular interest is the case in which $X$ is the circle.

\begin{defn}
Let $u(x,t)$ be a two--dimensional system defined on a circle $X$.
Its spatial complexification $Z(x)$ is assumed to be always non--zero
and continuous on $X$. Then $Z(x)$ can be written as
$Z(x)=C(x)e^{iQ(x)}$.
As $Z(0)=Z(2\pi)$ we have $Q(2\pi)=Q(0)+n2\pi$ where $n$ is an integer.
The number $n$ is called the {\bf spatial winding number} of the system.
\end{defn}

%%%%%%%%%%%%%%%%%%%%%% MODULATIONS

Let us now introduce the notion of modulation of a system :

\begin{defn}
Let $u(x,t)$ be a two--dimensional system which BOD is given by
\ref{bod2d}. Let ${\cal M}=(M(x),N(t))$ be a (regular) modulatrix.
A spatiotemporal (regular) {\bf modulation} of $u(x,t)$ is the system
$u'(x,t)$ defined by

\begin{equation}
u'(x,t)=\alpha'_1\psi'_1(t)\phi'_1(x)+\alpha'_2\psi'_2(t)\phi'_2(x)
\end{equation}

with

\begin{equation}
\alpha'_1\phi'_1(x)+i\alpha'_2\phi'_2(x)=M(x)(\alpha_1\phi_1(x)+
i\alpha_2\phi_2(x))
\end{equation}

and

\begin{equation}
\alpha'_1\psi'_1(t)+i\alpha'_2\psi'_2(t)=N(t)
(\alpha_1\psi_1(t)+i\alpha_2\psi_2(t)).
\end{equation}

\end{defn}

Note that in general $\psi'_1$, $\phi'_1$, $\psi'_2$, and $\phi'_2$
are not eigenvectors of $u'$.

\medskip

Let us investigate the effect of a modulation on the dimension of
the system.

\begin{thm}\label{thmdim}
Let $u(x,t)$ be a system described by two phase continuous
complexifications $Z(x)=C(x)e^{iQ(x)}$ and $Y(t)=D(t)e^{iR(t)}$.
Let ${\cal M}=(M,N)$ be a continuous phase modulatrix with
$M(x)= A(x)e^{iF(x)}$ and $N(t)=B(t)e^{iG(t)}$. The dimension of
the modulated system is reduced to one if

\begin{equation}
\forall x \in X, \ Q(x)+F(x)=z_1
\end{equation}

or

\begin{equation}
\forall t \in T, \ R(t)+G(t)=z_2
\label{condredt}
\end{equation}

where $z_1$ and $z_2$ are two complex numbers. Otherwise the
dimension of the modulated system remains equal to two.
\end{thm}

\begin{pf}
Let us first  apply the spatial modulation.
Let $Z'(x)$ be the spatial complexification of the spatialy modulated
system $u'(x,t)$ :

\begin{equation}
Z'(x)=M(x)Z(x)
\end{equation}

The dimension of the spatial structure is reduced only if it
 is embedded in a segment, i.e., if $Z'$ can be written as
$Z'(x)=C'(x)e^{iq_1}$ where $q_1$ is real.
In this case,

\begin{equation}
u'(x,t)=C'(x)\cos q_1 \psi_1(t)+C'(x)\sin q_1 \psi_2(t),
\end{equation}

it is clear that the new system is one--dimensional,
more precisely

\begin{equation}
u'(x,t)=\alpha'_1\psi'_1(t)\phi'_1(x)
\end{equation}

with $\alpha'_1=||C'||$, $\phi_1=C'/\alpha'_1$, and $\psi'_1=
\cos q_1 \psi_1+\sin q_1 \psi_2$.

\medskip

However,

\begin{equation}
C'(x)e^{iq_1}=A(x)e^{iF(x)}C(x)e^{iQ(x)}
\end{equation}

This equation implies $\forall x \in X,\ Q(x)+F(x)=e^{iq_1}$, where
$z_1=e^{iq_1}$. Conversely, if we assume that $M(x)=A(x)e^{i(q_1-F(x))}$,
then the modulated system becomes one--dimensional.
We can show in the same way that the reduction of the dimension
of the dynamics after a temporal modulation is equivalent to
(\ref{condredt}).\qed\end{pf}

\begin{thm}
A regular spatial modulation does not change the spatial winding number.
\end{thm}

\begin{pf}
Let $Z(x)=C(x)e^{iQ(x)}$ be a complexification with a winding number $n$.
We then have $Q(2\pi)=Q(0)+n2\pi$.
Let $M(x)=A(x)e^{iF(x)}$ be a regular spatial modulation. Therefore,
because $X$ is here  the circle, $F(2\pi)=F(0)$.
The spatial complexification of the modulated system is by definition

\begin{equation}
Z'(x)=C'(x)e^{iQ'(x)}
\end{equation}

with $C'(x)=C(x)A(x)$ and $Q'(x)=Q(x)+F(x)$.
Then using the values of $F(x)$ and $Q(x)$ at $x=0$ and $x=2\pi$
we get $Q'(2\pi)=Q'(0)+n2\pi$.\qed\end{pf}

Note that this theorem implies that in order to change the winding number
by modulation, a non--regular modulation is required.

\begin{thm}
Let $u(x,t)$  be a two--dimensional system defined on the circle $X$
 whose spatial complexification $Z(x)$ and its temporal
complexification $Y(t)$ never vanish.

Then there exists one unique regular modulatrix and a pair
consisting of an
integer $k$ and a real $\omega$ such that $u(x,t)$ is the modulation
of a monochromatic travelling wave
$u_0(x,t)=cos(kx)cos(\omega t)+sin(kx)sin(\omega t)$.\label{theomod}
\end{thm}

\begin{pf}
Let us write the spatial complexification $Z(x)$ as

\begin{equation}
Z(x)=C(x) e^{iQ(x)}
\end{equation}

and the temporal complexification as

\begin{equation}
Y(t)=D(t)e^{iR(t)}
\end{equation}

We set

\begin{equation}
k=\frac{1}{2\pi}D_Q
\end{equation}

and

\begin{equation}
\omega=\frac{1}{t_1-t_0}D_R,
\end{equation}

where $D_Q=Q(2\pi)-Q(0)$ and $D_R=R(t_1)-R(t_0)$ are the increases
 of the arguments of $Z$ and $Y$ respectively.
The numbers $k$ and $\omega$ are unique,
because $D_Q$ and $D_R$ are unique.
Because $X$ is a circle, the winding number $k$ is an integer.

\medskip

Let us define :

\begin{equation}
M(x)=C(x)e^{i[Q(x)-kx]}
\end{equation}

and

\begin{equation}
N(t)=D(t)e^{i[R(t)-\omega t]}
\end{equation}

The modulatrix ${\cal M}=(M,N)$ is regular and $u(x,t)$ is the
 modulation of a system $u_0(x,t)$ by ${\cal M}$.
 The spatial complexification and the temporal complexification
 of $u_0(x,t)$ are respectively $Z_0(x)=e^{ikx}$ and
$Y_0(t)=e^{i\omega t}$. The system $u_0(x,t)$ is thus of the form

\begin{equation}
u_0(x,t)=cos(kx)cos(\omega t)+sin(kx)sin(\omega t).
\end{equation}

The functions $Q(x)$ and $R(t)$ are defined modulo $2\pi$, so $C(x)$
 and $D(t)$ are unique.
Thus the functions $M$ and $N$ are unique.
\qed\end{pf}

This theorem provides a way to consider a two--dimensional experimental
system as a spatio--temporal modulation of a monochromatic
travelling wave.
In particular, a continuous deformation of a wave
is described by a modulatrix ${\cal M_\epsilon}=(M_\epsilon,N_\epsilon)$
that depends on the control parameter $\epsilon$.
In the case where $X$ is the circle, we can define the winding number,
and the jump in the winding number that is observed for a certain
value of the control parameter will correspond to a bifurcation.
Note the analogy of the complexifications with the loops in the
order parameter space in the study of the topological
defects \cite {Mermin}.

\medskip

Our complex modulation is a generalisation of the real modulation
introduced in \cite{LimaMod}.
The modulations introduced in \cite{LimaMod} are not appropriate
for the description of the deformations of a function $\phi_1(x)$
 (resp. $\psi_1(t)$) that has a zero which shifts as $\epsilon$ varies.
Instead, we can describe the deformation of such a function provided
that we find a complementary function $\phi_2(x)$
(resp. $\psi_2(t)$)that never vanishes for the value of $x$
(resp. $t$) where $\phi_1(x)$ (resp. $\psi_1(t)$) has a zero.
Note that in the absence of resonances, as in the case of real
modulatrix ${\cal M}=(M,N)$,
the new eigenvectors are
the modulation of the non perturbed system, as in Ref.~\cite{LimaMod}.

 \section{Identifying the modulated monochromatic travelling waves}
\label{LookModTW}
%%%%%%%%%%%%%%%%%%%%%%%%%%%%%%%%%%%%%%%%%%%%%

The aim of this section is to simplify the description of the
dynamical behaviour of the present system by  the  identification
of the  two--dimensional structures existing in the system for
 different values of the control parameter $\epsilon$.
In order to identify two--dimensional subsystems whose spatial and
temporal complexification never vanishes, the idea that this system is
a deformed monochromatic travelling wave (MTW) is used.
More precisely, a MTW has two basic properties : (i) the degeneracy
of the weights and (ii)the Fourier transform of the chronos and
topos are delta functions.

\medskip

We show in Fig.~\ref{poitot} the plot of the weight on a logarithmic
scale.
In the analysis of the plot of weights, we have to look for pairs of
eigenvalues that are degenerated.
The degeneracy of pairs is obvious only for the first two eigenvalues
of each data set.
However, if the ratio $a_n/a_{n+1}$ is plotted versus $n$,
the degeneracy of all weights can be quantified (see Fig.~\ref{poitot})

\def\textplotpoic{Plot of the weights for the data in y-log scale (left)
 and plot of the ratios $a_{n+1}/a_n$ with respect to $n$ (right).
(a) data ${\cal U}_1$ (b) data ${\cal U}_2$(c) data ${\cal U}_3$.
They exhibit the degeneracy of the weights which may be associated
to a spatio-temporal symmetry}

\begin{figure}
%\centering
%\begin{tabular}[t]{c}
%\subfigure[]
%{\label{poi1}\epsfig{file={plotpoitot1.eps},width=\textwidth}}\\
%\subfigure[]
%{\label{poi2}\epsfig{file={plotpoitot2.eps},width=\textwidth}}\\
%\subfigure[]
%{\label{poi3}\epsfig{file={plotpoitot3.eps},width=\textwidth}}
%\end{tabular}
\caption{\textplotpoic}
\label{poitot}
\end{figure}

We are thus looking for values of $n$ such that $a_n/a_{n+1}$
is close to unity.
This plot provides a way to couple eigenvalues :
two eigenvalues $a_n$ and $a_{n+1}$ are considered to be coupled
if the ratio $a_n/a_{n+1}$ forms a local maximum in the plot  ratios.
Following the previous rule, we obtain the pairs
listed in Table.~\ref{tabpairpoi} that we call structures.

\begin{table}[htb]
 \begin{center}
  \caption{Pairs obtained after the analysis of the weigths}
  \label{tabpairpoi}
  \begin{tabular}{l|c|c|c|c|c|c|c|c}
 %  data set &    &    &  &    &     &      &      &      \\
  data set       \\
\hline
${\cal U}_1$ &1--2&3--4& -- &6--7&9--10&12--13&14--15&16--17\\
${\cal U}_2$ &1--2&3--4& -- &7--8&9--10&12--13&14--15&16--17\\
${\cal U}_3$ &1--2& -- &6--7&8--9&10--11&12--13&14--15&16--17\\
\hline
  \end{tabular}
 \end{center}
\end{table}
%

%*************************
Notice that, considering  only the weight distribution is not
sometimes sufficient to decide the pairing. Such is, for instance,
the case in data ${\cal U}_3$ for the $a_n$, $n=3,4,5$. A careful look
to the corresponding chronos and topos, specially by considering
their Fourier transform, will give, in this case an
unambiguous pairing.
%*************************

The next step in the identification of modulated MTW is to consider pairs
of functions that have the same spatial (temporal) frequencies.
An appropriate way to detect such modulated sine and cosine functions
is to apply a Fourier transform to the eigenfunctions.
The plot of the square modulus of the Fourier transforms of the
chronos and topos are represented in Fig.~\ref{bodfftc} and
Fig.~\ref{bodfftt}.

\begin{figure}
\centering
\caption{Modulus squared Fourier transforms of the chronos.
The three columns correspond to the three data sets.
The $x$--axis is the frequency. The frequency unit is
$d\omega=2048.10^{-6}s^{-1}$. The $y$ axis is labelled
by the eigenvector index. }
\label{bodfftc}
\end{figure}

\begin{figure}
\centering
\caption{Modulus squared Fourier transforms of the topos.
The three columns correspond to the three data sets.
The $x$--axis is the spatial frequency. The frequency unit is
the spatial loop. The $y$ axis is labelled
by the eigenvector index.}
\label{bodfftt}
\end{figure}

A single--peaked spectrum is found for the first two chronos and topos
of each data set. Pairs of eigenvectors are found by  inspection :
each two Fourier spectra with almost or exactely the same
structure indicates the presence of a pair. Using the conditions
on the Fourier transform, Table~\ref{tabpairpoi} can be improved.
Table ~\ref{tabpairtf} give the new list of pairs. The
pairs are labelled $w_k$, where $k$ is the index of the wave associated
with the pair of eigenvectors.

\begin{table}[htb]
 \begin{center}
  \caption{Pairs obtained after the analysis of the Fourier spectrum
 of the eigenvectors. The modulated MTW associated with those pairs
are labeled $w_k$, where $k$ is the index of the wave.}
  \label{tabpairtf}
  \begin{tabular}{l|c|c|c|c|c|c|c|c}
 %  data set &    &    &  &    &     &      &      &      \\
  data set   &$w_1$&$w_2$&$w_3$&$w_4$&$w_5$&$w_6$&$w_7$&$w_8$\\
\hline
${\cal U}_1$ &1--2&3--4&5--6&7--8&9--10&12--13&14--15&16--17\\
${\cal U}_2$ &1--2&3--4&5--6&7--8&9--10&12--13&14--15&16--17\\
${\cal U}_3$ &1--2&3--4&5--6&7--8&9--10&12--13&14--15&16--17\\
\hline
  \end{tabular}
 \end{center}
\end{table}

Note that the Eigenvector 11 for the sets ${\cal U}_1$ and
${\cal U}_2$  and Eigenvector 7 for the set ${\cal U}_3$
do not belong to a pair.

%***************
Note also that the degeneracy of the weights $a_6$ and $a_7$
doesn't correspond to a modulated travelling wave. This is because
the energy of the eigenfunction 7 wich is a global oscillation
of the plasma without any propagation is close to the energy
of the third travelling wave.
%***************

\section{\label{secorg}Organization of the different modulated MTWs}
%%%%%%%%%%%%%%%%%%
In the previous section, two--dimensional subsystems
which are modulated travelling waves were identified .
In this section, these waves
are studied with respect to both the eigenvector indexes and the
control parameter.

\medskip

First the spatial and temporal frequencies $k$ and $\omega$ are
determined. Actually, Theorem \ref{theomod} provides an explicit way
to compute $k$ and $\omega$. However, this direct approach is
cumbersome in practice. Therefore the winding number $k$ is obtained
in a different manner : The first eight spatial
structures associated with the first eight modulated travelling waves
are shown in Fig.~\ref{strucx}.
The winding number becomes easier to inspect in the plot of the spatial
structures (as in the case of the Fourier transform) as the control
parameter becomes larger. This shows that the spatial modulation
becomes more regular as $\epsilon$ increases.

\begin{figure}
\centering
\caption{Spatial structures associated with the first eight modulated
travelling waves. a) Data set ${\cal U}_1$,  b) Data set
${\cal U}_2$, c) Data set ${\cal U}_3$. Each square subfigure is
labelled in the corner
by the index of the eigenvectors spanning the structure.}
\label{strucx}
\end{figure}

Fig.~\ref{strucx} shows that the winding number for the data set
${\cal U}_1$ is well defined only for the first wave $w_1$ (The
labelling is defined in the Table ~\ref{tabpairtf}).
For the data set ${\cal U}_2$, the spatial winding number is defined
for the three first waves, and for the data set ${\cal U}_3$ it is well
defined for all eight waves studied. The winding numbers found
are shown in the Table ~\ref{tabkomeg}.
The temporal frequencies $\omega$ associated with these waves are
determined from the Fourier spectra (Fig.~\ref{bodfftc}).
A broad Fourier spectra corresponds
to the presence of zeros in the temporal complexifications.
The $\omega$ is well--defined from the Fourier spectra only for
the first wave of the data set
${\cal U}_1$ and ${\cal U}_2$, and for the first three waves for the
data set ${\cal U}_3$.

\begin{table}[htb]
 \begin{center}
  \caption{Spatial winding numbers $k$ and temporal frequencies $\omega$
associated with the first eight waves. The wave number $i$ is labelled
wave $w_i$. }
  \label{tabkomeg}
  \begin{tabular}{l|c|c|c|c|c|c|c|c|}
          &$w_1$&$w_2$&$w_3$&$w_4$&$w_5$&$w_6$&$w_7$&$w_8$\\ \hline
data set ${\cal U}_1$ &$k=2$      & -- & --& -- & --& -- & -- & --\\
             &$\omega=20$& -- & --& -- & --& -- & -- & --\\ \hline
data set ${\cal U}_2$ &$k=2$      &$k=3$ & $k=3$ & -- & --& -- &
 -- & --\\
             &$\omega=22$& -- & --& -- & --& -- & -- & --\\ \hline
data set ${\cal U}_3$ &$k=2$      &$k=1$ & $k=3$&$k=4$&$k=5$ &
 $k=6$&$k=7$ & $k=8$ \\
             &$\omega=44$&$\omega=25$&$\omega=63$& -- & --& -- &
-- & --\\ \hline
  \end{tabular}
 \end{center}
\end{table}

The phase speed of a modulated wave is defined as the ratio of the
temporal frequency $\omega$ to the spatial frequency $k$ of the
associated unmodulated monochromatic wave.
Considering the first waves given in Tab.~\ref{tabkomeg}
it is noted that  a speed doubling occurs when the control
parameter $\epsilon$ increases from the value that belongs to
the data set ${\cal U}_2$ to that of the data set ${\cal U}_3$.
Indeed, the phase speed associated with the waves $w_1$, $w_2$, and $w_3$
in the data set ${\cal U}_3$ is very close to twice the speed of the
first travelling wave of the data ${\cal U}_2$.
In the next section,
a simple model for this spatio--temporal bifurcation is discussed.

%%%%%%%%%%%%%%%%%%%%%%%%%%%

 Another way to study the effect of the spatial modulation is to plot
a graylevel plot of the two dimensional restriction $u_{m,n}(x,t)$
associated with the modulated wave. The graylevel plots for the
first eight waves of each data set is presented in
Fig.~\ref{2Dgraytot}.

\begin{figure}
\centering
\caption{Graylevel plots of the two--dimensional structures.
Each column is associated with a data set. For each column, the letters a)
\dots h) indicate the wave $w_1$,\dots,$w_8$ that is plotted. For each
figure, the $x$--axis is time and the $y$--axis space.}
\label{2Dgraytot}
\end{figure}

In those plots a phase spatial modulation implies a distortion
of the wave front, this distorsion depending only on $x$,
and an amplitude spatial modulation corresponding to global crashes
in the amplitude of the waves, those crashes having the same intensity
for fixed positions. In the graylevel plot, the phase modulation
is easier to observe than in the probe structure, where a phase
modulation implies only a non-uniform spacing of the vectors $\xi_x$.

\medskip

The graylevel plots show that in the data sets ${\cal U}_1$ and
${\cal U}_2$, the first wave has a phase defect that localy makes
the wave front more vertical. The other waves are strongly phase
modulated with strong tearing of the wave fronts.
However, the wave fronts of the waves in the data set ${\cal U}_2$
are more  regular than in the data set ${\cal U}_1$.
In the data set ${\cal U}_3$,
all the strong phase defects have disappeared or been reduced.

\medskip

In the gray-level
plots the strong amplitude modulations can be observed as well.
The global  crashes of the amplitude of the wave correspond to
an amplitude modulation close to zero. This amplitude modulation
is itself chaotic.

%%%%%%%%%%%%%%%%%%%%%%%%%%%%%%%%%

The data set ${\cal U}_3$ is thus  a state where  chaos is
essentially present in time.
The spatial structures are, on the contrary, fairly regular.
The temporal chaos is then due to a chaotic modulation acting  on a
non--chaotic spatio--temporal structure, i.e. a travelling wave.
Note that, furthermore, in this case we are faced to an
homogeneous turbulence \cite{Lumley} as it is seen in Fig.~\ref{bodfftc}.
Also note that the chaotic modulation of monochromatic
waves has a close relationship with
the three--wave interaction model of  the drift wave instability
\cite{horton90}.
However, it one may ask if the restriction to three waves
is pertinent here.
Indeed, our study reveals a set of at least eight
waves relevant to the data set ${\cal U}_3$.

\medskip

It is easier to study the spatial amplitude modulation
in the probe structure plots [Fig.\ref{strucx}].
The modulation of the first wave increases with $\epsilon$.
In contrast, the next waves become more regular as $\epsilon$ increases.
The spatial structure of the data set ${\cal U}_3$ coincides
with a large scale of eigenvectors ($m=3,\dots,19$).

\medskip

To study these large scale phenomena, the plot of the weights is
considered [Fig.~\ref{poitot}]. Neglecting the pairwise degeneracy,
 the weights $a_n$ decrease exponentially
with $n$ in well defined regions.
In particular, in the ditribution of the weights for the data set
${\cal U}_3$, the boundaries of such a region are given by the indices
$n_1=7$ and $n_2=20$. These boundaries correspond to a pronounced
broadening of the Fourier spectrum in the low--frequency regime
[cf. Fig.~\ref{bodfftc}]. In this region, the spatial structures
have a well defined winding number as shown in
Fig.~\ref{strucx}.
Note that the spectrum of the chronos is bounded by a frequency
near 60 in the data set ${\cal U}_1$ and ${\cal U}_2$.
In the data set ${\cal U}_3$ this bound has been shifted to 250.
Even in the eigenvectors with a broad spectrum, i.e. those
with an index greater than 23, the frequencies higher than 250
are nonexistent.
However, the spatial frequencies are not bounded and increase with the
index of the eigenvector.

%******************
\begin{rem}
Even if the radial dependency of the electric field is not
directly accessible to the present diagnostic, we may wonder
that such effects is present giving rise to a collective rotation
of the whole plasma column and therefore to a shift in the (Fourier)
dispersion relation for the waves.
This effect seems present in the frequencies reported in Table 4,
for the data ${\cal U}_3$. However, since we restrict ourselves
to each two dimensional structures , this fact would have no effect
in our analysis. It simply changes the position of $\chi(T)$ in
the space of the functions of time $H(X)$ by a global rotation. We thank
one of the referees to have pointed us this question.
\end{rem}
%********************

\section{\label{secmodel}Model for the speed doubling}
%%%%%%%%%
In the previous section, it was noted that a speed doubling occurs when
the control parameter increases.
The data set ${\cal U}_2$ corresponded to a state
before this bifurcation and the data set ${\cal U}_3$
corresponded to a state after the bifurcation.
In this section we give a simple model for this speed doubling.
Our model is built with two two--dimensional structures.

\medskip

We first describe one of the two structures.
The figure Fig.~\ref{modelsketch} shows how a spatial modulation
can change the winding number and the speed of the speed of the
eigenstructure.

\begin{figure}
\centering
\caption{A spatial modulation can imply a topological change in
the spatial structure, which corresponds to a change
of the speed of the modulated travelling wave. The spatial structures,
the trajectories, and a sketch of the crest of the wave are plotted for
four values of the control parameter.}
\label{modelsketch}
\end{figure}

The phase speed (cf. Section~\ref{secorg}) is defined as a ratio.
In order to define the states to model uniquely, the speed
alone is not sufficient.
The model structure describes an evolution from a state described by the
function $u_{-1}(x,t)=\cos(2x)\cos(\omega t)+\sin(2x)\sin(\omega t)$
to a state described by the function
$u_{1}(x,t)=\cos(x)\cos(\omega t)+\sin(x)\sin(\omega t)$.

\medskip

Let us consider a temporal complexification
which is  independant on $\epsilon$ defined by

\begin{equation}
Y_0(t)=e^{i\omega t}.
\end{equation}

and the corresponding spatial complexification by

\begin{equation}
Z_{\epsilon}(x)=(1-\epsilon)e^{i2x}+(1+\epsilon)e^{ix}.
\end{equation}

These spatial  and temporal complexifications correspond for
$\epsilon=-1$ to the complexifications of  $u_{-1}(x,t)$
and  for $\epsilon=1$ to the complexification of
 $u_{1}(x,t)$.
Fig.~\ref{model} shows the plots for four values
of the control parameter $\epsilon$,
the spatial complexification $Z_{\epsilon}(x)$ in the right column,
and the gray level representation of the two dimensionnal
system $u_{\epsilon}(x,t)$.

\begin{figure}
\centering
\caption{Model for the speed doubling. The gray level plot of
the modulated MTW is shown for four values of $\epsilon$.
The corresponding spatial structures are shown in the right column.}
\label{model}
\end{figure}

The spatial complexification is never equal to zero except
for the value $\epsilon=0$.
For this value, the spatial complexification vanishes for
$x=\pi$.
For negative $\epsilon$ the winding number is 2, and for positive
$\epsilon$ it is 1.
However, for $\epsilon=0$, the winding number is not defined.
A bifurcation in the winding number thus occurs.

\medskip

A regular spatial modulation is just defined (i) before the bifurcation
by dividing the spatial complexification $Z_{\epsilon}(x)$ by
$Z_{-1}(x)$ (ii) after the bifurcation by dividing the spatial
complexification $Z_{\epsilon}(x)$ by $Z_{1}(x)$.
The spatial modulation before the bifurcation $M^b_{\epsilon}(x)$
is defined by

\begin{equation}
M^b_{\epsilon}(x)=(1-\epsilon)+(1+\epsilon)e^{-ix}.
\end{equation}

$M^b_{\epsilon}(x)$ is a regular modulation because
$(1-\epsilon)>(1+\epsilon)$ when $\epsilon<0$.
The spatial modulation after the bifurcation $M^a_{\epsilon}(x)$
is defined by

\begin{equation}
M^a_{\epsilon}(x)=(1+\epsilon)+(1-\epsilon)e^{ix}
\end{equation}

$M^a_{\epsilon}(x)$ is a regular modulation after the bifurcation
because $(1+\epsilon)>(1-\epsilon)$ when $\epsilon<0$.
The spatial defect which occurs at $x=\pi$ for
$\epsilon=0$ permits the wave front to change its
shape as shown in  Fig.~\ref{model}.
The two (unnormalized) topos are the real part and the imaginary
part of the spatial complexification

\begin{equation}
\phi_1(x)=(1-\epsilon)\cos(2x)+(1+\epsilon)\cos(x),
\end{equation}

\begin{equation}
\phi_2(x)=(1-\epsilon)\sin(2x)+(1+\epsilon)\sin(x).
\end{equation}

The chronos are left unchanged

\begin{equation}
\psi_1(t)=\cos(\omega t),
\end{equation}

\begin{equation}
\psi_2(x)=\sin(\omega t).
\end{equation}

Both the spatial behaviour (see Fig~\ref{strucx},
Fig~\ref{2Dgraytot})---i.e. the bifurcation of a winding number
from the value 2 to the value 1--- and the temporal behaviour
(see  Fig~\ref{bodfftc}, Fig~\ref{2Dgraytot})---i.e.
a frequency $\omega$ roughly independant of the control parameter
$\epsilon$ shows that this structure corresponds to the modulated MTW
$w_1$ for the data sets ${\cal U}_1$ and ${\cal U}_2$ and
$w_2$ for the data sets ${\cal U}_3$.
(The notations $w_k$ were introduced in the Table ~\ref{tabpairtf}).

\medskip

In the same way we model the structure corresponding to the
second pair in ${\cal U}_1$ and ${\cal U}_2$ and the first in
${\cal U}_3$. Notice that in this case the bifurcation is due to
the simultaneous deformation of the temporal modulation and of
the corresponding spatial modulation.

\medskip

Finally, the model is built by glueing the two structures weighted
by two corresponding eigenvalues (the relative energies of the
two structures). According to the general theory of bifurcations
described by the BOD \cite{LimaSym}, these energies cross
at the bifurcation and this is the reason why the energies of
the two structures are interchanged when passing from ${\cal U}_2$
to ${\cal U}_3$.\section{Summary and Conclusions}
%%%%%%%%%%%%%%%%%%%
The BOD of the experimental data of the drift--waves experiment
showed the importance of the two--dimensional sub--systems.
It has been shown that the two--dimensional structures are
complex--modulated monochromatic travelling waves (modulated MTW).
A spatial regularity counterpart of a temporal chaos has been
discovered in the most turbulent data set.
The most important feature of the evolution of the
system with the control parameter $\epsilon$, i.e. the speed
doubling, has been modeled as well.
The model consists of a pair of simple two--dimensional structures that
undergo an exchange of energy as  the value of the
control parameter varies.

\medskip

A study of new experimental data will be done in the future
in order to improve the model and better characterize the bifurcation.
The regular behaviour also needs to be better understood.
Furthermore, the connection between the modulated MTW and the
waves--interactions models for the drift--waves will be
investigate in the future.

{\bf Acknowledgments}

One of us (A.M.) kindly aknowledges Prof A. Piel of the Institut
f\"ur Experimentalphysik at Kiel for his warm hospitality
during part of this work. A.M. also thanks very much Prof. R. Lima.
This work would not have been possible without him.
We also thank Dr. T. Dudok de Wit for fruitful discussions.
Thanks also
to Dr. P. N. Watts who helped to improve the manuscript.

\end{document}